\documentstyle[12pt]{article}

\newcommand{\be}{\begin{equation}}
\newcommand{\ee}{\end{equation}}
\newcommand{\bea}{\begin{eqnarray}}
\newcommand{\eea}{\end{eqnarray}}
\newcommand{\sptwo}{1.8}
\newcommand{\doublespace}{\edef\baselinestretch{\sptwo}\Large\normalsize}
\newcommand{\newsection}[1]{
\section{#1}
\setcounter{equation}{0}}

\newcounter{newapp}
\begin{document}
\vspace{1.0in}
\begin{center}
{\large\bf Goldstino Couplings to Matter}
\end{center}
\begin{center}
T.E. Clark and S.T. Love\\
Department of Physics\\
Purdue University\\
West Lafayette, IN 47907-1396
\end{center}

\vspace{1.0in}
\begin{center}
{\bf Abstract}
\end{center}

A nonlinearly realized supersymmetric action describing the 
invariant couplings of the Goldstino to matter is constructed. 
Using the Akulov-Volkov Lagrangian, any operator can be made part of 
a supersymmetric invariant action. Of particular interest are interaction 
terms which include the product of the Akulov-Volkov Lagrangian with the 
ordinary matter Lagrangian as well as the coupling of the product of the 
covariant derivative of the Goldstino field to the 
matter supersymmetry current. The later is the lowest 
dimensional operator linear in the 
Goldstino field. A Goldstino Goldberger-Treiman relation 
is established and shown to be satisfied by the effective action.

\pagebreak

\doublespace

\newsection{Introduction}

If supersymmetry (SUSY) is to be realized in nature, it must be as a broken 
symmetry. The breaking mechanism which maintains the perponderance 
of the dynamical constraints of the symmetry and hence is theoretically most 
attractive, is a spontaneous one. Indeed many of the currently investigated 
attempts to construct realistic models of electroweak symmetry breaking 
incorporating SUSY use spontaneous 
supersymmetry breaking in one form or another. This includes both the 
so-called hidden sector $\cite{hid}$ and visible sector classes of 
models $\cite{vis}$. 

A general consequence of the spontaneous 
breakdown of global supersymmetry is the 
appearance of a Nambu-Goldstone fermion, 
the Goldstino $\cite{AV},~\cite{FI}$. The leading term in the  
action describing its self dynamics at energy 
scales below $\frac{4\pi}{\kappa}$, where $\frac{1}{\kappa}$ is the Goldstino 
decay constant, is uniquely fixed by the Akulov-Volkov 
effective Lagrangian $\cite{AV}$ which takes the form
\be
\label{LAV}
{\cal{L}}_{AV}=-\frac{1}{2\kappa^2}~\det{A}
\ee
where 
\be
A_\mu~^\nu\equiv\delta_\mu^\nu+i\kappa^2\lambda\stackrel{\leftrightarrow}
{\partial}_\mu\sigma^\nu\bar{\lambda}~~.
\ee
Here $\lambda (\bar{\lambda})$ is the Goldstino Weyl spinor field. 
This effective Lagrangian provides a valid description of the Goldstino self 
interactions independent of the particular (non-perturbative) 
mechanism by which the SUSY is 
dynamically broken $\cite{DSB}$. 
Moreover, if the spontaneously broken 
supersymmetry is gauged, with the erstwhile Goldstino degrees of freedom  
absorbed to become the londitudinal (spin 1/2) modes of the gravitino 
via the super-Higgs mechanism, then 
the action formed from the Akulov-Volkov 
Lagrangian also describes the dynamics of those modes. 
This is completely analogous to using the gauged non-linear sigma model to 
represent the dynamics of the longitudinal degrees of freedom of the 
$W_\pm$ and $Z_0$ vector bosons independent of the particular 
mechanism employed to 
break the electroweak symmetry.

	Nonlinear realizations of symmetry transformations 
allow a model independent  analysis of the dynamical
consequences of spontaneous symmetry breaking using the
Nambu-Goldstone degrees of freedom.  This is true 
whether the part of the theory responsible for the symmetry 
breaking is strongly
or weakly interacting.  If weakly interacting, such as in the 
standard electroweak model  with a light Higgs scalar, 
then the low energy physics can be directly calculated in
perturbation theory.  On the other hand, if the symmetry breaking sector of the 
model is strongly interacting, then explicit direct calculations
of dynamical consequences can prove quite difficult.  Since there are many
such models of SUSY breaking presently studied,  it is worthwhile to 
determine, in a completely model independent way, the various consequences
of the supersymmetry breaking.  

Using non-linear realizations 
of supersymmetry for both the Goldstino and non-Goldstino degrees of freedom,  
Samuel and Wess $\cite{W}$ constructed supersymmetric invariant couplings of 
the Goldstino to matter. Their construction entailed a somewhat 
elaborate procedure in which the Goldstino field and all matter 
fields are promoted to become superfields whose lowest components are the 
ordinary fields themselves and whose higher components involve the 
product of Goldstino fields and derivatives of the lowest components. For the 
special case of the Goldstino promoted superfields, the $\theta$ or 
$\bar{\theta}$ components also 
contain the Goldstino decay constant as an additive component. Using these 
superfields, every ordinary operator can then be cast as part of a 
manifestly supersymmetric action. While this procedure is elegant and complete, 
it does require the introduction of a considerable amount of additional 
(super) structure. 

In this paper, we present an alternate construction of a 
non-linearly realized supersymmetric invariant action. Our procedure works 
directly with the ordinary (component) Goldstino and matter fields and does 
not require the 
introduction of entire superfield structures. Thus in a simple and 
straightforward manner, we can make any ordinary operator part of a 
manifestly supersymmetric action. After introducing the non-linear SUSY 
transformations and covariant derivative, we construct 
the SUSY (and internal symmetry) 
invariant action terms using the special properties of the  
Akulov-Volkov Lagrangian. We focus on two particular interaction terms. One 
involves the coupling of the Akulov-Volkov Lagrangian to the ordinary 
matter action. Once the ordinary matter action is appropriately normalized, 
the coefficient of this term is fixed solely by the Goldstino decay constant. 
Another action term, which is the lowest dimensional operator linear in the 
Goldstino field, involves the coupling of its (SUSY covariant) derivative 
to the ordinary matter supersymmetry current. This coupling is then used to 
show that a Goldberger-Treiman relation $\cite{GT}$ associated with the 
spontaneous supersymmetry breaking is indeed satisfied.

\pagebreak

\newsection{Nonlinear SUSY Transformations}

The self dynamics of the Goldstino can be encapsulated in the Akulov-Volkov 
Lagrangian Eq.~($\ref{LAV}$). 
The supersymmetry transformations are nonlinearly realized on the Goldstino 
field by
\bea
\label{SRG}
\delta^Q(\xi,\bar{\xi})\lambda^\alpha&=&\frac{1}{\kappa}\xi^\alpha-i\kappa
(\lambda\sigma^\rho\bar{\xi}-\xi\sigma^\rho\bar{\lambda})\partial_\rho
\lambda^\alpha\cr
\delta^Q(\xi,\bar{\xi})\bar{\lambda}_{\dot\alpha}&=&\frac{1}{\kappa}
\bar{\xi}_{\dot\alpha}-i\kappa
(\lambda\sigma^\rho\bar{\xi}-\xi\sigma^\rho\bar{\lambda})\partial_\rho
\bar{\lambda}_{\dot\alpha}~~,
\eea
where $\xi^\alpha,\bar{\xi}_{\dot\alpha}$ are Weyl spinor SUSY transformation 
parameters. The Akulov-Volkov Lagrangian then transforms as a total divergence
\be
\delta^Q(\xi,\bar{\xi}){\cal{L}}_{AV}=-i\kappa\partial_\rho
[(\lambda\sigma^\rho\bar{\xi}-\xi\sigma^\rho\bar{\lambda}){\cal{L}}_{AV}]
\ee
and hence the associated action
\be
I_{AV}=\int d^4x ~{\cal{L}}_{AV}
\ee
is SUSY invariant:
\be
\delta^Q(\xi,\bar{\xi})I_{AV}=0~~.
\ee

The supersymmetry algebra can also be nonlinearly realized on the matter 
(non-Goldstino) fields, generically denoted by $\phi_i$, where $i$ can 
represent any Lorentz or internal symmetry labels, as
\be
\label{SSR}
\delta^Q(\xi,\bar{\xi})\phi_i=-i\kappa
(\lambda\sigma^\rho\bar{\xi}-\xi\sigma^\rho\bar{\lambda})
\partial_\rho\phi_i~~.
\ee
This is referred to as the standard realization $\cite{W}$. 
Forming the SUSY Ward identity functional differential operator
\bea
\delta^Q(\xi,\bar{\xi})&=&\int d^4x[\delta^Q(\xi,\bar{\xi})\lambda^\alpha
\frac{\delta}{\delta\lambda^\alpha}+\delta^Q(\xi,\bar{\xi})
\bar{\lambda}_{\dot\alpha}
\frac{\delta}{\delta \bar{\lambda}_{\dot\alpha}} \cr
&&~~~~+\sum_{i}\delta^Q(\xi,\bar{\xi})
\phi_i\frac{\delta}{\delta\phi_i}],
\eea
one readily establishes the SUSY algebra
\bea
 ~ [\delta^Q(\xi,\bar{\xi}),\delta^Q(\eta,\bar{\eta})]&=&
 -2i\delta^{P}(\xi\sigma\bar{\eta}-\eta\sigma\bar{\xi})\cr
 ~ [\delta^Q(\xi,\bar{\xi}),\delta^{P}(a)]&=&0~~.
\eea
As usual, the space-time translations are given by
\bea
\delta^P(a)\lambda^\alpha&=&a^\mu\partial_\mu\lambda^\alpha\cr
\delta^P(a)\bar{\lambda}_{\alpha}&=&
a^\mu\partial_\mu\bar{\lambda}_{\alpha}\cr
\delta^P(a)\phi_i &=&a^\mu\partial_\mu\phi_i~~,
\eea
with $a^\mu$ the global space-time translation parameter and 
\be
\delta^P(a)=\int d^4x [\delta^P(a)\lambda^\alpha\frac{\delta}
{\delta\lambda^\alpha}+\delta^P(a)\bar{\lambda}_{\dot\alpha}
\frac{\delta}{\delta 
\bar{\lambda}_{\dot\alpha}}
+\sum_{i}\delta^P(a)\phi_i\frac{\delta}{\delta\phi_i}]
\ee
is the space-time translation Ward identity function differential operator. 

Under the non-linear SUSY standard realization, 
the derivative of a matter field transforms as
\bea
\label{SSRD}
\delta^Q(\xi,\bar{\xi})(\partial_\nu\phi_i)&=&\partial_\nu
[\delta^Q(\xi,\bar{\xi})\phi_i]\cr
&=&-i\kappa(\lambda\sigma^\rho\bar{\xi}-\xi\sigma^\rho\bar{\lambda})
\partial_\rho(\partial_\nu\phi_i)
-i\kappa\partial_\nu(\lambda\sigma^\rho\bar{\xi}-\xi\sigma^\rho\bar
{\lambda})
\partial_\rho\phi_i ~~.\cr
&&
\eea
In order to eliminate the second term on the RHS and thus 
restore the SUSY covariance, we introduce a SUSY covariant derivative 
which transforms analogously to $\phi_i$. To achieve this, we note that
\be
\delta^Q(\xi,\bar{\xi})A_\mu~^\nu=-i\kappa[(\lambda\sigma^\rho\bar{\xi}-\xi
\sigma^\rho\bar{\lambda})\partial_\rho A_\mu~^\nu+\partial_\mu
(\lambda\sigma^\rho
\bar{\xi}-\xi\sigma^\rho\bar{\lambda})A_\rho~^\nu]~~,
\ee
from which it follows that 
\bea
\delta^Q(\xi,\bar{\xi})(A^{-1})_\mu~^\nu&=&-(A^{-1})_\mu~^\rho
[\delta^Q(\xi,\bar{\xi})A_\rho~^\sigma](A^{-1})_\sigma~^\nu\cr
&=&-i\kappa[(\lambda\sigma^\rho\bar{\xi}-\xi
\sigma^\rho\bar{\lambda})\partial_\rho(A^{-1})_\mu~^\nu
\cr
&&~~-\partial_\rho(\lambda\sigma^\nu\bar{\xi}-\xi
\sigma^\nu\bar{\lambda})(A^{-1})_\mu~^\rho]~~,
\eea
where 
\be
(A^{-1})_\mu~^\nu A_\nu~^\rho=\delta_\mu^\rho~~.
\ee
We are thus led to define the non-linearly realized SUSY covariant derivative as
\be
\label{SCD}
{\cal{D}}_\mu\phi_i=(A^{-1})_\mu~^\nu\partial_\nu\phi_i~~,
\ee
so that under the standard realization of SUSY:
\be
\delta^Q(\xi,\bar{\xi})({\cal{D}}_\mu\phi_i)=-i\kappa
(\lambda\sigma^\rho\bar{\xi}-\xi\sigma^\rho\bar{\lambda})\partial_\rho
({\cal{D}}_\mu\phi_i)~~.
\ee

In addition to the SUSY and space-time translations, we 
can also define R-transformations under which the Goldstino field transforms 
as $\cite{CL}$
\bea
\delta^R(\omega)\lambda^\alpha&=&i\omega\lambda^\alpha\cr
\delta^R(\omega)\bar{\lambda}_{\dot\alpha}&=&-i\omega
\bar{\lambda}_{\dot\alpha}~~.
\eea
Forming the R-transformation Ward identity functional differential operator
\be
\delta^R(\omega)=\int d^4x[\delta^R(\omega)\lambda^\alpha\frac{\delta}{\delta
\lambda^\alpha}+\delta^R(\omega)\bar{\lambda}_{\dot\alpha}
\frac{\delta}{\delta\bar{\lambda}_{\dot\alpha}}+\sum_{i}\delta^R(\omega)\phi_i
\frac{\delta}{\delta\phi_i}]~~,
\ee
it is readily established that the algebra
\be
[\delta^R(\omega),\delta^Q(\xi,\bar{\xi})]=
\delta^Q(-i\omega\xi,i\omega\bar{\xi})
\ee
holds independent of form of $\delta^R(\omega)\phi_i$. The action formed from 
the Akulov-Volkov Lagrangian is invariant under R-symmetry, supersymmetry and 
space-time translations. Moreover the improved currents associated with these 
symmetries have been shown $\cite{CL}$ to form the components of 
a supercurrent $\cite{FZ}$. 
Thus all conservation laws and anomalies are derivable from the supercurrent 
conservation law and the generalized trace anomaly $\cite{CPS}\cite{GMZ}$.

Since the Goldstino field transforms as a singlet under any internal 
symmetry transformation, $\delta^G(\Lambda)\lambda^\alpha=0=\delta^G(\Lambda)
\bar{\lambda}_{\dot\alpha}$, 
the Akulov-Volkov action is also invariant under 
internal symmetry transformations:
\be
\delta^G(\Lambda)I_{AV}=0~~,
\ee
where $\Lambda$ parametrizes the transformation. 
Denoting the internal symmetry matter field transformation as 
$\delta^G(\Lambda)\phi_i$, 
then the Ward identity functional differential operator 
characterizing the internal symmetry transformation is
\be
\delta^G(\Lambda)=\int d^4x \sum_{i}\delta^G(\Lambda)\phi_i
\frac{\delta}{\delta\phi_i}
\ee
Note that if the internal symmetry is gauged, the non-linearly realized 
SUSY, gauge covariant
derivative, Eq.~($\ref{SCD}$), is replaced with  
\be
{\cal{D}}_\mu\phi_i=(A^{-1})_\mu~^\nu D_\nu\phi_i~~,
\ee
where $D_\mu\phi_i$ is the ordinary gauge covariant derivative.

\pagebreak

\newsection{Invariant Actions}

We now construct actions containing the Goldstino and matter fields 
which are invariant under both SUSY and internal symmetry transformations. 
The Akulov-Volkov action is one such term which contains the Goldstino 
kinetic term and self couplings. Using the Akulov-Volkov Lagrangian, we can 
form SUSY invariant actions out of any Lorentz and 
internal symmetry singlet operator 
${\cal{O}}={\cal{O}}(\phi,{\cal{D}}\phi)$. To achieve this, we note that 
under the non-linear standard realization of SUSY given by Eqs.~($\ref{SRG},
\ref{SSR},\ref{SSRD}$), such an operator transforms as 
\be
\delta^Q(\xi,\bar{\xi}){\cal{O}}(\phi,{\cal{D}}\phi)=-i\kappa
(\lambda\sigma^\rho\bar{\xi}-\xi\sigma^\rho\bar{\lambda})\partial_\rho
{\cal{O}}(\phi,{\cal{D}}\phi)~~.
\ee
Consequently the action
\bea
I_{{\cal{O}}}&=&-2\kappa^2 C_{{\cal{O}}}\int d^4x ~{\cal{L}}_{AV}~{\cal{O}}
\cr
&=&C_{{\cal{O}}}\int d^4x ~(\det{A}) ~{\cal{O}}~~,
\eea
with $C_{{\cal{O}}}$ a constant, is SUSY invariant:
\be
\delta^Q(\xi,\bar{\xi})I_{{\cal{O}}}=0~~.
\ee
Since ${\cal{L}}_{AV}$ is defined so as to contain the additive constant term 
$-\frac{1}{2\kappa^2}$ (or equivalently, the $\det{A}$ starts with the 
identity), the action $I_{{\cal{O}}}$ includes the piece 
$C_{{\cal{O}}}\int d^4x {\cal{O}}(x)$ for any operator ${\cal{O}}$. 

One special case is afforded by using the 
internal symmetry invariant 
ordinary matter Lagrangian ${\cal{L}}_\phi(\phi,{\cal{D}}\phi)$ where 
all derivatives are replaced by SUSY covariant 
derivatives. Under SUSY
\be
\delta^Q(\xi,\bar{\xi}){\cal{L}}_{\phi}(\phi,{\cal{D}}\phi)=-i\kappa
(\lambda\sigma^\rho\bar{\xi}-\xi\sigma^\rho\bar{\lambda})\partial_\rho
{\cal{L}}_{\phi}(\phi,{\cal{D}}\phi)~~,
\ee
while under the internal group transformation, the Lagrangian is invariant:
\be
\delta^G(\Lambda){\cal{L}}_{\phi}(\phi,{\cal{D}}\phi)=0~~.
\ee
It follows that the action
\be
I_{LL}=-2\kappa^2\int d^4x ~{\cal{L}}_{\phi}(\phi,{\cal{D}}\phi)~{\cal{L}}_{AV}
(\lambda,\bar{\lambda})
\ee
is both SUSY and internal symmetry invariant
\be
\delta^Q(\xi,\bar{\xi})I_{LL}=0
\ee
\be
\delta^G(\Lambda)I_{LL}=0~~.
\ee
Note that in the absence of Goldstino fields, this action reduces to the 
ordinary matter action $I_\phi=\int d^4x 
{\cal{L}}_{\phi}(\phi,\partial_\mu\phi)$ so $I_{LL}$ 
contains the ordinary matter action as well as couplings of the Goldstino to 
matter. Thus once the normalization of the ordinary matter action is fixed, 
so are its couplings to the Goldstino field. 
As such this term requires no additional independent coupling constant.
Further note that using the non-linear realization, various higher dimensional
operators such as the electron anomalous magnetic moment operator, can also
be made part of a SUSY invariant action.  On the other hand, such a term
cannot be included in a SUSY invariant action if the supersymmetry is linearly
realized $\cite{FR}$.

Both $I_{AV}$ and $I_{LL}$ depend on $\lambda,\bar{\lambda}$ only through 
$A^{\mu\nu}$ and thus only through the bilinear combination 
$\lambda \bar{\lambda}$ (and derivatives). While, by using the Goldstino 
field, any Lorentz and internal symmetry singlet can incorporated into a 
supersymmetric invariant action, the most natural setting is to consider 
those pure matter actions which allow for linear  
realizations of the supersymmetry. In that case, using the 
associated internal symmetry 
singlet supersymmetry currents $Q_{\phi ~\alpha}^\mu(\phi,\partial_\mu\phi)$ 
and $\bar{Q}_{\phi ~\dot\alpha}^\mu(\phi,\partial_\mu\phi)$, 
we can construct another 
invariant action whose Goldstino dependence is odd in 
$\lambda, \bar{\lambda}$ and in fact 
starts off as linear in $\partial_\mu\lambda$. 
Letting $Q_{\phi ~\alpha}^\mu=Q_{\phi ~\alpha}^\mu(\phi,{\cal{D}}\phi)$
and $\bar{Q}_{\phi ~\dot\alpha}^\mu=\bar{Q}_{\phi ~\dot\alpha}^\mu
(\phi,{\cal{D}}\phi)$ be the matter supercurrents 
where all space-time derivatives are replaced by non-linearly 
realized SUSY covariant derivatives, it follows 
that under the standard realization of SUSY that
\bea
\delta^Q(\xi,\bar{\xi})Q_{\phi ~\alpha}^\mu&=&-i\kappa
(\lambda\sigma^\rho\bar{\xi}-\xi\sigma^\rho\bar{\lambda})\partial_\rho 
Q_{\phi ~\alpha}^\mu\cr
\delta^Q(\xi,\bar{\xi})\bar{Q}_{\phi ~\dot\alpha}^\mu&=&-i\kappa
(\lambda\sigma^\rho\bar{\xi}-\xi\sigma^\rho\bar{\lambda})\partial_\rho 
\bar{Q}_{\phi ~\dot\alpha}^\mu
\eea
while
\be
\delta^G(\Lambda)Q_{\phi ~\alpha}^\mu=0=\delta^G(\Lambda)
\bar{Q}_{\phi ~\dot\alpha}^\mu~~.
\ee
When used in conjunction with the SUSY transformations:
\bea
\delta^Q(\xi,\bar{\xi})({\cal{D}}_\mu\lambda^\alpha)&=&-i\kappa
(\lambda\sigma^\rho\bar{\xi}-\xi\sigma^\rho\bar{\lambda})\partial_\rho
({\cal{D}}_\mu\lambda^\alpha)\cr
\delta^Q(\xi,\bar{\xi})({\cal{D}}_\mu\bar{\lambda}_{\dot\alpha})&=&-i\kappa
(\lambda\sigma^\rho\bar{\xi}-\xi\sigma^\rho\bar{\lambda})\partial_\rho
({\cal{D}}_\mu\bar{\lambda}_{\dot\alpha})
\eea
we construct the invariant action
\be
I_{\lambda Q}=-2\kappa^3 C_Q\int d^4x 
~{\cal{L}}_{AV}~({\cal{D}}_\mu\lambda^\alpha 
Q_{\phi ~\alpha}^\mu+\bar{Q}_{\phi ~\dot\alpha}^\mu {\cal{D}}_\mu\bar{\lambda}
^{\dot\alpha})
\ee
where $C_Q$ is a constant. This action satisfies
\be
\delta^Q(\xi,\bar{\xi})I_{\lambda Q}=0
\ee
\be
\delta^G(\Lambda)I_{\lambda Q}=0~~.
\ee
Using the form of the Akulov-Volkov Lagrangian and the SUSY covariant 
derivative, we see that
\be 
I_{\lambda Q}=\kappa C_Q\int d^4x
[\partial_\mu\lambda^\alpha Q_{\phi ~\alpha}^\mu
(\phi,\partial_\nu\phi)+\bar{Q}_{\phi ~\dot\alpha}^\mu(\phi,\partial_\nu\phi)
\partial_\mu\bar{\lambda}^{\dot\alpha}]+...
\ee
which is coupling linear in the Goldstino field. In fact, this mass dimension 6 
operator contains the smallest power of $\kappa$ coefficient of the 
various couplings of the Goldstino to matter. The appearance of the coupling 
of the Goldstino field to the divergence of the matter 
supersymmetry current is 
certainly anticipated. In fact, it is reminiscent of the situation in 
spontaneously broken chiral symmetry where the Nambu-Goldstone pion couples 
derivatively to the spontaneously broken matter chiral symmetry current.

Combining the various terms, we secure the SUSY and internal symmetry invariant 
action
\bea
\label{EFFACT}
I&=&I_{AV}+I_{LL}+I_{\lambda Q}\cr
&=&\int d^4x ~{\cal{L}}_{AV}-2\kappa^2\int d^4x ~{\cal{L}}_{AV}~{\cal{L}}_{\phi}
\cr
&&~~~~-2\kappa^3 C_Q\int d^4x ~{\cal{L}}_{AV}~({\cal{D}}_\mu\lambda^\alpha 
Q_{\phi ~\alpha}^\mu+\bar{Q}_{\phi ~\dot\alpha}^\mu {\cal{D}}_\mu\bar{\lambda}
^{\dot\alpha})
\eea
which satisfies
\be
\delta^Q(\xi,\bar{\xi})I=0
\ee
and
\be
\delta^G(\Lambda)I=0~~.
\ee
The action starts out as
\be
\label{GSCC}
I=\int d^4x [{\cal{L}}_{AV}+{\cal{L}}_{\phi}+\kappa C_Q 
(\partial_\mu\lambda^\alpha Q_{\phi ~\alpha}^\mu
+\bar{Q}_{\phi ~\dot\alpha}^\mu\partial_\mu\bar{\lambda}^{\dot\alpha})+...]
\ee
so that  
\bea
\frac{\delta I}{\delta \lambda^\alpha}&=&\frac{\delta I_{AV}}
{\delta \lambda^\alpha}-\kappa C_Q 
\partial_\mu Q_{\phi ~\alpha}^\mu+...\cr
&=&-\frac{i}{2}\det{A} (A^{-1})^{\nu\mu}(\sigma_\nu\partial_\mu
\bar{\lambda})_\alpha-\kappa C_Q 
\partial_\mu Q_{\phi ~\alpha}^\mu+...\cr
&=&-i\sigma^\mu_{\alpha{\dot\alpha}}\partial_\mu\bar{\lambda}^{\dot\alpha}
-\kappa C_Q \partial_\mu Q_{\phi ~\alpha}^\mu+...
\eea
and the Goldstino field equation takes the form
\be
\label{GFE}
\sigma^\mu_{\alpha{\dot\alpha}}\frac{1}{i}
\partial_\mu\bar{\lambda}^{\dot\alpha}=\kappa 
C_Q\partial_\mu Q_{\phi ~\alpha}^\mu+...~~~~.
\ee
\pagebreak

\newsection{Goldberger-Treiman Relation}

As a consequence of their Nambu-Goldstone nature, the coupling of the 
Goldstino to matter is constrained to satisfy certain general relationships. 
One such constaint is the analog of the Goldberger-Treiman 
relationship $\cite{GT}$ familiar from pion physics and spontaneously 
broken chiral symmetry. 
When applied to spontaneously broken supersymmetry, the analogous relation 
ties form factors of the supersymmetry 
and Goldstino currents at zero momentum transfer to the Goldstino decay 
constant and mass differences between matter boson and fermion states. 
The Lorentz decomposition of the 
supersymmetry current $Q_\alpha^\mu$ taken between arbitrary single particle 
(scalar) Bose and (spin 1/2) 
Fermi states, $|p_1;B>$  and $|p_2;F>$, of masses $m_B$ and $m_F$ and 
carrying 4-momenta 
$p_1^\mu$ and $p_2^\mu$, respectively, takes the form
\bea
\label{LDSC}
&&<p_1;B|Q_\alpha^\mu(0)|p_2;F>=[A_1(q^2)q^\mu+A_2(q^2)k^\mu+A_3(q^2)\sigma^\mu
\bar{\sigma}\cdot q]_\alpha~^\beta\chi_\beta(p_2)_F\cr
&&~~~~~~+[A_4(q^2)\sigma^\mu+A_5(q^2)q^\mu\sigma \cdot q
+A_6(q^2)k^\mu\sigma \cdot q]
_{\alpha {\dot\alpha}}\bar{\chi}^{\dot\alpha}(p_2)_F\cr
&&
\eea 
where $q^\mu=(p_1-p_2)^\mu$ and $k^\mu=(p_1+p_2)^\mu$ and the 
fermion spinors satisfy
\bea
\sigma\cdot p_2\bar{\chi}(p_2)_F&=&-m_F\chi(p_2)_F\cr
\bar{\sigma}\cdot p_2\chi(p_2)_F&=&-m_F\bar{\chi}(p_2)_F~~.
\eea
Conservation of the supersymmetry current, $\partial_\mu Q^\mu_\alpha=0$ then 
relates the various form factors as
\be
\label{CSC}
q^2[A_1(q^2)-A_3(q^2)]=(m_B^2-m_F^2)A_2(q^2)~~.
\ee

Since the massless Goldstino directly couples to the supersymmetry current, 
some of these form factors are singular in the $q^2\rightarrow 0$ limit. 
Thus before taking this limit, we need to include the effect of 
the massless Goldstino pole. This pole 
is reflected in the non-vanishing 
matrix element of the supersymmetry current between the 
vacuum and single Goldstino state, $|q;\lambda>$, of 4-momentum $q^\mu$ which 
is given by
\be
<0|Q_\alpha^\mu(0)|q;\lambda>=\frac{1}{i\kappa}\sigma^\mu_{\alpha{\dot\alpha}}
\bar{\chi}^{\dot\alpha}_\lambda~~,
\ee
where $\kappa^{-1}$ is the Goldstino decay constant.  

It follows that the combination $Q^\mu_\alpha-
\frac{1}{i\kappa}\sigma^\mu_{\alpha{\dot\alpha}}\bar{\lambda}^{{\dot\alpha}}$ 
has vanishing matrix element between the vacuum and single Goldstino state. 
The matrix element of this 
combination taken between the single Bose state and single Fermion state 
can then be Lorentz decomposed just as in Eq.~($\ref{LDSC}$) where now the 
various form factors are all non-singular in the $q^2\rightarrow 0$ limit. 

Finally, the Goldstino current $j^G_\alpha$ is defined 
through the Goldstino field equation 
\be
\sigma^\mu_{\alpha{\dot\alpha}}\frac{1}{i}
\partial_\mu\bar{\lambda}^{\dot\alpha}=
j^G_\alpha~~.
\ee
Taking its matrix element between the Bose and Fermi states leads to the 
Lorentz decomposition 
\be
<p_1;B|j^G_\alpha(0)|p_2;F>=B_1(q^2)\chi_\alpha(p_2)_F+B_2(q^2)(\sigma\cdot q)
_{\alpha{\dot\alpha}}\bar{\chi}^{\dot\alpha}(p_2)_F
\ee
and thus 
\be
\label{GLD}
<p_1;B|\bar{\lambda}^{\dot\alpha}(0)|p_2;F>=-\frac{B_1(q^2)}{q^2}
(\bar{\sigma}\cdot q)^{{\dot\alpha}\alpha}\chi_\alpha(p_2)_F+B_2(q^2)
\bar{\chi}^{\dot\alpha}(p_2)_F~~.
\ee
Since the form factors of the combination $Q^\mu_\alpha-
\frac{1}{i\kappa}\sigma^\mu_{\alpha{\dot\alpha}}\bar{\lambda}^{{\dot\alpha}}$ 
are regular as $q^2\rightarrow 0$, we see on comparison of Eqs.~($\ref{LDSC}$) 
and ($\ref{GLD}$) that the $A_1(q^2)$ form factor is regular 
while the $A_3(q^2)$ form factor is singular. The singular piece is given by  
\be
\lim_{q^2\rightarrow 0}q^2A_3(q^2) = \frac{i}{\kappa}B_1(0)~~.
\ee
Sustituting into Eq.~($\ref{CSC}$) and taking the $q^2\rightarrow 0$ limit we 
secure the Goldstino Goldberger-Treiman relation
\be
\label{GTR}
-\frac{i}{\kappa}B_1(0)=(m_B^2-m_F^2)A_2(0)~~.
\ee

To establish that the effective action Eq.~($\ref{EFFACT}$) satisfies 
this Goldstino Goldberger-Treiman relation, we note that a Noether 
construction of the conserved supersymmetry current starts out as
\be
Q^\mu_\alpha=C_Q Q^\mu_{\phi ~\alpha}+...~~,
\ee
while the Goldstino field 
equation Eq.~($\ref{GFE}$) provides the identification of the Goldstino 
current as
\be
j^G_\alpha=\kappa C_Q\partial_\mu Q^\mu_{\phi ~\alpha}+...~~~~.
\ee
For the ``matter" supersymmetry current we use $\cite{FZ}$ $\cite{DWF}$
\be
Q^\mu_{\phi ~\alpha}=\partial^\mu \bar{A}\psi_\alpha+...~~~~,
\ee
where $\bar{A} ~(\psi)$ are the Bose (Fermi) fields creating (destroying) the 
Bose (Fermi) states in the matrix elements of the Lorentz decomposition of the 
supersymmetry and Goldstino currents. The matrix elements are readily computed 
yielding
\bea
A_2(0)&=&-\frac{i}{2}C_Q\cr
B_1(0)&=&\frac{\kappa}{2}(m_B^2-m_F^2)C_Q
\eea
and the Goldberger-Treiman relation, Eq.~($\ref{GTR}$) is indeed satisfied.

This work was supported in part by the U.S. Department of Energy under 
grant DE-FG02-91ER40681 (Task B).

\pagebreak

\newpage
\end{document}